\documentclass[aps,pra,twocolumn,floatfix,twoside]{revtex4}
\usepackage{amssymb}
\usepackage{graphicx}
\usepackage{dcolumn}
\usepackage{bm}

\begin{document}

\title{Density expectation value of two independent interacting
Bose-Einstein condensates}
\author{Hongwei Xiong$^{1,2,3}$, Shujuan Liu$^{1,2,3}$, Mingsheng Zhan$%
^{1,2} $}

\address{$^{1}$State Key Laboratory of Magnetic Resonance and
Atomic and Molecular Physics, Wuhan Institute of Physics and
Mathematics, Chinese Academy of Sciences, Wuhan 430071, P. R.
China}
\address{$^{2}$Center for Cold Atom Physics, Chinese Academy of Sciences, Wuhan 430071, P.
R. China}
\address{$^{3}$Graduate School of the Chinese Academy of
Sciences, P. R. China}

\date{\today }

\begin{abstract}
After removing the double-well potential trapping two initially independent
Bose condensates, the density expectation value is calculated when both the
exchange symmetry of identical bosons and interatomic interaction are
considered. After the overlapping, it is shown that there is a nonzero
interference term in the density expectation value. This nonzero
interference term physically arises from the exchange symmetry of identical
bosons and interatomic interaction which make two initially independent
condensates become coherent after the overlapping. It is found that the
calculated density expectation value with this model agrees with the
interference pattern observed in the experiment by Andrews \textit{et al }
\textit{(Science \textbf{275}, 637 (1997))}.

\end{abstract}

\maketitle

The coherence property plays an essential role in the wave nature
of Bose-Einstein condensates (BECs), and it has been investigated
intensively \cite{Meystre} after the experimental realizations of
BECs in dilute gases. For two coherently separated Bose
condensates, it is not surprising that there is a clear
interference pattern when the two condensates are allowed
to overlap. In the celebrated experiment by Andrews \textit{et al }\cite%
{Andrew}, however, high-contrast fringes were observed even for two
completely independent condensates at an initial time. This experimental
result shows clearly that there is a coherence property after the
overlapping of two independent interacting condensates.

In many literature (see for example \cite{Leggett,Pethick} and references
therein), upon expansion, it is shown that there is no interference term in
the density expectation value for two initially independent condensates, and
thus the observed high-contrast fringes \cite{Andrew} for two independent
condensates were interpreted \cite{Leggett} with the aid of the high-order
correlation function $P\left( \mathbf{r},\mathbf{r}^{\prime },t\right) $
(which is an oscillation function of $\mathbf{r}-\mathbf{r}^{\prime }$) and
quantum measurement process. The contradiction between the vanishing
interference term in the density expectation value and the observed
high-contrast fringes was also discussed based on several theories such as
the stochastic simulations of the photon detection for atoms \cite{JAV}, the
expansion of Fock state by the linear superposition of coherent states \cite%
{CASTIN}, and the continuous measurement theory \cite{Zoller}. In addition,
the nonlinear effects in the interference pattern of two coherently
separated Bose condensates were investigated in Refs \cite{Rohrl,Liu}.

In the present work, we calculate carefully the density expectation value
for two initially independent condensates by including the interatomic
interaction. It is found that \textit{for the case of two initially
independent condensates, upon expansion, there is a nonzero interference
term in the density expectation value when interatomic interaction and the
exchange symmetry of identical bosons are taken into account carefully!}
After removing the double-well potential trapping the two initially
independent condensates, we give the theoretical result of the density
expectation value which agrees with the interference pattern observed in
\cite{Andrew}.

For the condensate trapped in a magnetic trap, it can be divided into two
condensates by a blue-detuned laser beam. The combined double-well potential
takes the form%
\begin{eqnarray}
V_{ext} &=&\frac{1}{2}m\left[ \omega _{x}^{2}\left( x-x_{0}\right)
^{2}+\omega _{y}^{2}y^{2}+\omega _{z}^{2}z^{2}\right]  \nonumber \\
&&+U_{0}e^{-\left( x-x_{0}\right) ^{2}/w_{x}^{2}-\left( y^{2}+z^{2}\right)
/w_{\perp }^{2}},  \label{external
potential}
\end{eqnarray}%
where the second term represents the external potential due to the laser
beam.

After the evaporative cooling for Bose gases trapped in the double-well
potential, if the intensity of the blue-detuned laser beam is sufficiently
low so that the tunneling effect is obvious, the two condensates can be
regarded to be coherently separated. For two coherently separated
condensates, every atom is described by the following normalization
wavefunction
\begin{equation}
\phi _{c}\left( \mathbf{r},t\right) =\left[ \sqrt{N_{1}}\phi _{1}\left(
\mathbf{r},t\right) +\sqrt{N_{2}}\phi _{2}\left( \mathbf{r},t\right) \right]
/\sqrt{N},
\end{equation}%
where $\phi _{1}\left( \mathbf{r},t\right) $ and $\phi _{2}\left( \mathbf{r}%
,t\right) $ are the normalization wave functions accounting for the two
condensates. $N=N_{1}+N_{2}$ is the total number of particles. Before the
overlapping of the two condensates, the average particle numbers in each
condensate are $N_{1}$ and $N_{2}$, respectively. After removing the
double-well potential, the density expectation value is then
\begin{eqnarray}
n_{c}\left( \mathbf{r},t\right) &=&N\left[ a_{c}\left\vert \phi _{1}\left(
\mathbf{r},t\right) \right\vert ^{2}+2b_{c}\times \mathrm{Re}\left( \phi
_{1}^{\ast }\left( \mathbf{r},t\right) \phi _{2}\left( \mathbf{r},t\right)
\right) \right.  \nonumber \\
&&\left. +c_{c}\left\vert \phi _{2}\left( \mathbf{r},t\right) \right\vert
^{2}\right] ,  \label{coherent-density}
\end{eqnarray}%
where $a_{c}=N_{1}/N$, $b_{c}=\sqrt{N_{1}N_{2}}/N$ and $c_{c}=N_{2}/N$. The
second term in the above equation accounts for the interference effect when
there is an overlapping between two condensates upon expansion.

If the intensity of the blue-detuned laser beam is sufficiently high so that
the tunneling effect can be omitted, the two condensates can be regarded to
be completely independent. In this case, the number of particles $N_{1}$ and
$N_{2}$ in each of the two condensates is fixed \cite{Pethick}. To
investigate clearly the role of the exchange symmetry of identical particles
and interatomic interaction, we calculate the density expectation value $%
n_{d}\left( \mathbf{r},t\right) $ directly from the many-body wave function.
Based on the well-known exchange symmetry of identical bosons \cite{Quam},
the many-body wavefunction of the whole system is given by%
\begin{eqnarray}
&&\Psi _{N_{1}N_{2}}\left( \mathbf{r}_{1},\mathbf{r}_{2},\cdots ,\mathbf{r}%
_{N},t\right)  \nonumber \\
&=&A_{n}\sqrt{\frac{N_{1}!N_{2}!}{N!}}\sum\limits_{P}P\left[ \phi _{1}\left(
\mathbf{r}_{1},t\right) \cdots \phi _{1}\left( \mathbf{r}_{N_{1}},t\right)
\times \right.  \nonumber \\
&&\left. \phi _{2}\left( \mathbf{r}_{N_{1}+1},t\right) \cdots \phi
_{2}\left( \mathbf{r}_{N_{1}+N_{2}},t\right) \right] ,  \label{wave-function}
\end{eqnarray}%
where $P$ denotes the $N!/(N_{1}!N_{2}!)$ permutations for the particles in
different single-particle state $\phi _{1}$ or $\phi _{2}$. The above
expression of the many-body wavefunction is based on the quantum mechanical
principle \cite{Quam}. $A_{n}$ is a normalization factor to assure $\int
\left\vert \Psi _{N_{1}N_{2}}\left( \mathbf{r}_{1},\mathbf{r}_{2},\cdots ,%
\mathbf{r}_{N},t\right) \right\vert ^{2}d\mathbf{r}_{1}d\mathbf{r}_{2}\cdots
d\mathbf{r}_{N}=1$. $A_{n}$ is determined by the following equation:
\begin{equation}
A_{n}\left[ \sum\limits_{i=0}^{\min \left( N_{1},N_{2}\right) }\frac{%
N_{1}!N_{2}!\left\vert \zeta \left( t\right) \right\vert ^{2i}}{i!i!\left(
N_{1}-i\right) !\left( N_{2}-i\right) !}\right] ^{1/2}=1,
\end{equation}%
where
\begin{equation}
\zeta \left( t\right) =\int \phi _{1}\left( \mathbf{r},t\right) \phi
_{2}^{\ast }\left( \mathbf{r},t\right) dV=\left\vert \zeta \left( t\right)
\right\vert e^{i\varphi _{c}}.  \label{xi}
\end{equation}%
In this Letter, to give a concise expression for various coefficients such
as $A_{n}$, we have introduced the rule $0^{0}=1$.

From the above many-body wavefunction, the exact density expectation value
takes the following form:
\begin{eqnarray}
n_{d}\left( \mathbf{r},t\right) &=&N\int \Psi _{N_{1}N_{2}}^{\ast }\left(
\mathbf{r},\mathbf{r}_{2},\cdots ,\mathbf{r}_{N},t\right) \times  \nonumber
\\
&&\Psi _{N_{1}N_{2}}\left( \mathbf{r},\mathbf{r}_{2},\cdots ,\mathbf{r}%
_{N},t\right) d\mathbf{r}_{2}\cdots d\mathbf{r}_{N}  \nonumber \\
&=&NB_{n}\left[ a_{d}\left\vert \phi _{1}\right\vert ^{2}+2b_{d}\times
\mathrm{Re}\left( e^{i\varphi _{c}}\phi _{1}^{\ast }\phi _{2}\right) \right.
\nonumber \\
&&\left. +c_{d}\left\vert \phi _{2}\right\vert ^{2}\right] ,  \label{density}
\end{eqnarray}%
where $B_{n}=A_{n}^{2}N_{1}!N_{2}!/N!$, and the coefficients in the above
equation are given by
\begin{eqnarray}
a_{d} &=&\sum\limits_{i=0}^{\min \left( N_{1}-1,N_{2}\right) }\frac{\left(
N-1\right) !\left\vert \zeta \left( t\right) \right\vert ^{2i}}{i!i!\left(
N_{1}-i-1\right) !\left( N_{2}-i\right) !},  \label{ad} \\
b_{d} &=&\sum\limits_{i=0}^{\min \left( N_{1}-1,N_{2}-1\right) }  \nonumber
\\
&&\frac{\left( N-1\right) !\left\vert \zeta \left( t\right) \right\vert
^{2i+1}}{i!\left( i+1\right) !\left( N_{1}-i-1\right) !\left(
N_{2}-i-1\right) !},  \label{bd} \\
c_{d} &=&\sum\limits_{i=0}^{\min \left( N_{1},N_{2}-1\right) }\frac{\left(
N-1\right) !\left\vert \zeta \left( t\right) \right\vert ^{2i}}{i!i!\left(
N_{1}-i\right) !\left( N_{2}-i-1\right) !}.  \label{cd}
\end{eqnarray}

For two independent ideal condensates, before the overlapping of the two
condensates, $\zeta \left( t=0\right) =0$. Based on the Schr\H{o}dinger
equation, it is easy to verify that after the double-well potential
separating the condensates is removed, one has $\zeta \left( t\right) =0$ at
any further time. Thus $b_{d}=0$, and the density expectation value is given
by%
\begin{equation}
n_{d}\left( \mathbf{r},t\right) =N_{1}\left\vert \phi _{1}\left( \mathbf{r}%
,t\right) \right\vert ^{2}+N_{2}\left\vert \phi _{2}\left( \mathbf{r}%
,t\right) \right\vert ^{2}.
\end{equation}
In this situation, the interference term is zero in the density expectation
value.

However, if the interatomic interaction is considered, $\zeta \left(
t\right) $ can be a nonzero value, and this will lead to a quite different
result compared with the ideal condensates. Assuming that $g=4\pi \hbar
^{2}a_{s}/m$ with $a_{s}$ being the scattering length, after removing the
double-well potential, the energy of the whole system is given by%
\[
E=\int \Psi _{N_{1}N_{2}}^{\ast }\left[ \sum_{i=1}^{N}-\frac{\hslash ^{2}}{2m%
}\nabla _{i}^{2}\right] \Psi _{N_{1}N_{2}}d\mathbf{r}_{1}\cdots d\mathbf{r}%
_{N}
\]
\begin{equation}
+g\int \Psi _{N_{1}N_{2}}^{\ast }\sum_{i<j}^{N}\delta \left( \mathbf{r}_{i}-%
\mathbf{r}_{j}\right) \Psi _{N_{1}N_{2}}d\mathbf{r}_{1}\cdots d\mathbf{r}%
_{N}.  \label{overallenergy}
\end{equation}%
The second term in the above equation represents the interaction energy of
the whole system by using the well-known two-body pseudopotentials.

By using the ordinary action principle and the energy of the whole system,
one can get the following coupled evolution equations for $\phi _{1}$ and $%
\phi _{2}$:
\begin{eqnarray}
i\hslash \frac{\partial \phi _{1}}{\partial t} &=&\frac{1}{N_{1}}\frac{%
\delta E}{\delta \phi _{1}^{\ast }},  \label{GP1} \\
i\hslash \frac{\partial \phi _{2}}{\partial t} &=&\frac{1}{N_{2}}\frac{%
\delta E}{\delta \phi _{2}^{\ast }},  \label{GP2}
\end{eqnarray}%
where $\delta E/\delta \phi _{1}^{\ast }$ and $\delta E/\delta \phi
_{2}^{\ast }$ are functional derivatives. With the above coupled equations,
one can understand that $\zeta \left( t\right) $ becomes nonzero after the
overlapping between two initially independent condensates for $g$ being a
nonzero value \cite{Xiong}.

Although generally speaking, $\left\vert \zeta \left( t\right) \right\vert $
is much smaller than $1$ because $\phi _{1}\phi _{2}^{\ast }$ is an
oscillation function about the space coordinate, nevertheless, a nonzero
value of $\zeta \left( t\right) $ will give significant contribution to the
density expectation value for large $N_{1}$ and $N_{2}$. In Fig. 1(a), we
give the relation between $b_{d}/c_{d}$ and $\zeta $ for $N_{1}=N_{2}=10^{3}$
based on the numerical calculations of Eqs. (\ref{ad}), (\ref{bd}) and (\ref%
{cd}). The relation between $b_{d}/c_{d}$ and $N_{1}=N_{2}$ for $\zeta
=0.001 $ is shown in Fig. 1(b). It is shown clearly that even for $%
\left\vert \zeta \right\vert $ being the order of $N_{1}^{-1}$, the value of
$b_{d}$ can not be omitted, and thus the interference term in Eq. (\ref%
{density}) will play an important role in the density expectation value.
Generally speaking, for $N_{1}\left\vert \zeta \right\vert >>1$ and $%
N_{2}\left\vert \zeta \right\vert >>1$, one has $b_{d}/\sqrt{a_{d}c_{d}}%
\approx 1$.

Generally speaking, increasing the particle number will enhance the effect
of the interference term in the density expectation value. Based on Eqs. (%
\ref{GP1}) and (\ref{GP2}), increasing the coupling constant $g$ will have
the effect of increasing $\zeta \left( t\right) $. Together with the
relation between $b_{d}/c_{d}$ and $\zeta $ illustrated in Fig. 1(a), this
shows that increasing the interatomic interaction will increase the effect
of the interference term.

\begin{figure}[tbp]
\includegraphics[width=0.8\linewidth,angle=270]{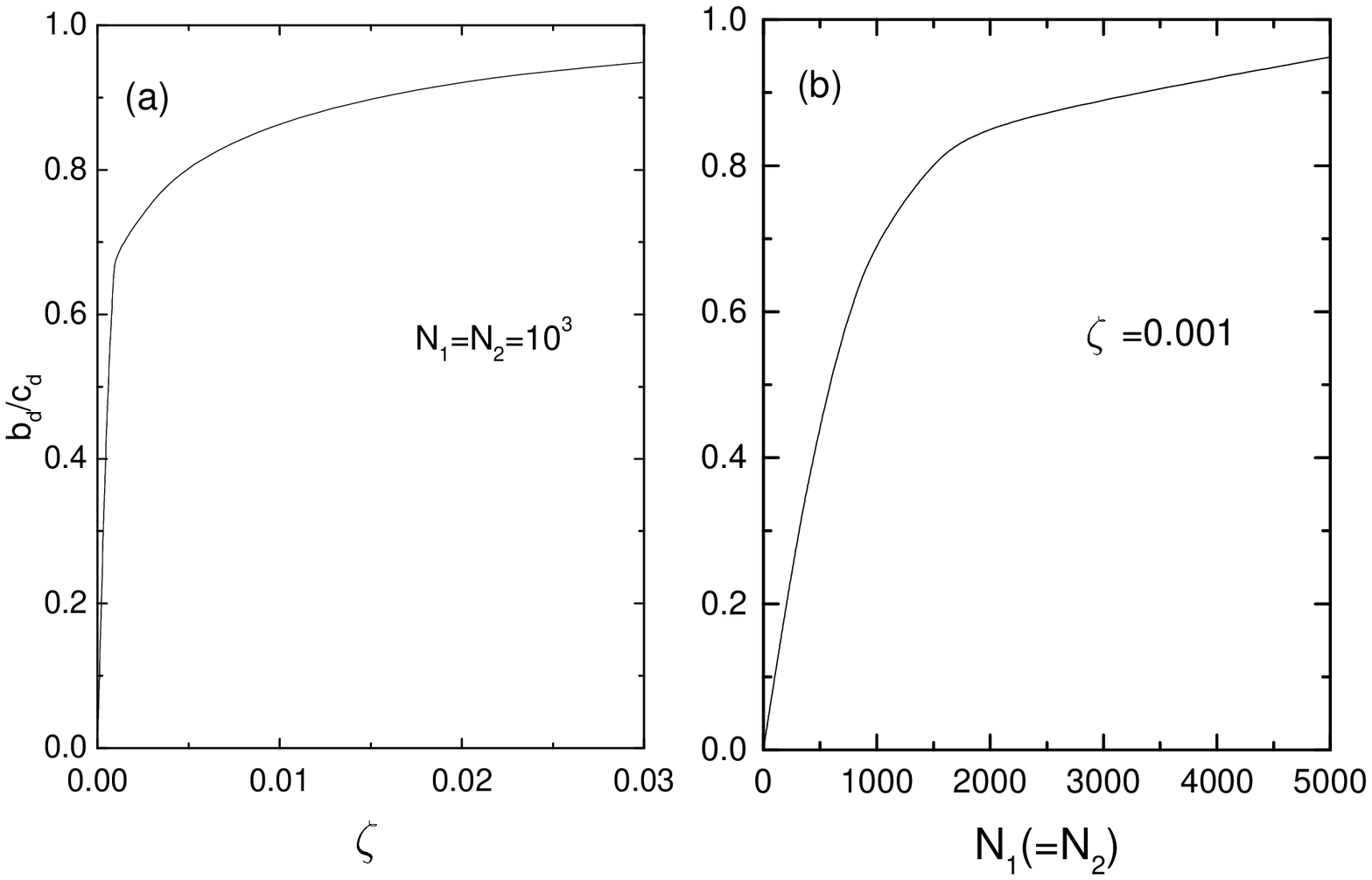}
\caption{Based on the numerical calculations of Eqs. (\protect\ref{ad}), (%
\protect\ref{bd}) and (\protect\ref{cd}), shown in Fig. 1(a) is the relation
between $b_{d}/c_{d}$ and $\protect\zeta $, while shown in Fig. 1(b) is the
relation between $b_{d}/c_{d}$ and $N_{1}=N_{2}$ for $\protect\zeta =0.001$.
It is shown clearly that the interference term can play an important role in
the density expectation value for $\left\vert \protect\zeta \right\vert $
being larger than $N_{1}^{-1}$.}
\end{figure}

In the experiment of Ref. \cite{Andrew}, $N=5\times 10^{6}$ condensed sodium
atoms were confined in a magnetic trap with $\omega _{x}=2\pi \times 18$
\textrm{Hz}, $\omega _{y}=\omega _{z}=2\pi \times 320$ \textrm{Hz}. A
blue-detuned laser beam of wavelength $514$ \textrm{nm} was focused into a
light sheet with a cross section of $12$ $\mathrm{\mu m}$ by $67$ $\mathrm{%
\mu m}$. The long axis of the laser beam was perpendicular to the long $x-$%
axis of the condensate. For a laser power of $14$ \textrm{mW}, the barrier
height is about $1.4$ $\mathrm{\mu K}$, which is much larger than the
chemical potential $\mu =0.03$ $\mathrm{\mu K}$. For this laser beam, the
two condensates can be regarded to be independent because they are well
separated and the tunneling effect can be omitted (See also \cite{Leggett}).
With these experimental parameters and the $s-$wave scattering length $%
a_{s}=2.75$ \textrm{nm}, the initial profile of the two condensates is shown
in Fig. 2. After the double-well potential is removed, the evolution of the
density expectation value $n_{d-x}\left( x,t\right) =\int n_{d}\left(
\mathbf{r},t\right) dydz$ (in unit of $N/2$) is given in Fig. 2 through the
numerical calculations of Eqs. (\ref{density}), (\ref{GP1}) and (\ref{GP2}).
We see that there is a clear interference pattern in the density expectation
value which agrees with the experimental result. Shown in the inset is the
evolution of $\left\vert \zeta \right\vert $ for these parameters. We have
also verified in the numerical calculation that $\left\vert \zeta
\right\vert $ is always zero for two ideal condensates. For the expansion
time of $40$ \textrm{ms}, the numerical result of $\left\vert \zeta
\right\vert $ shows that $b_{d}/c_{d}\approx 1$.

\begin{figure}[tbp]
\includegraphics[width=0.8\linewidth,angle=270]{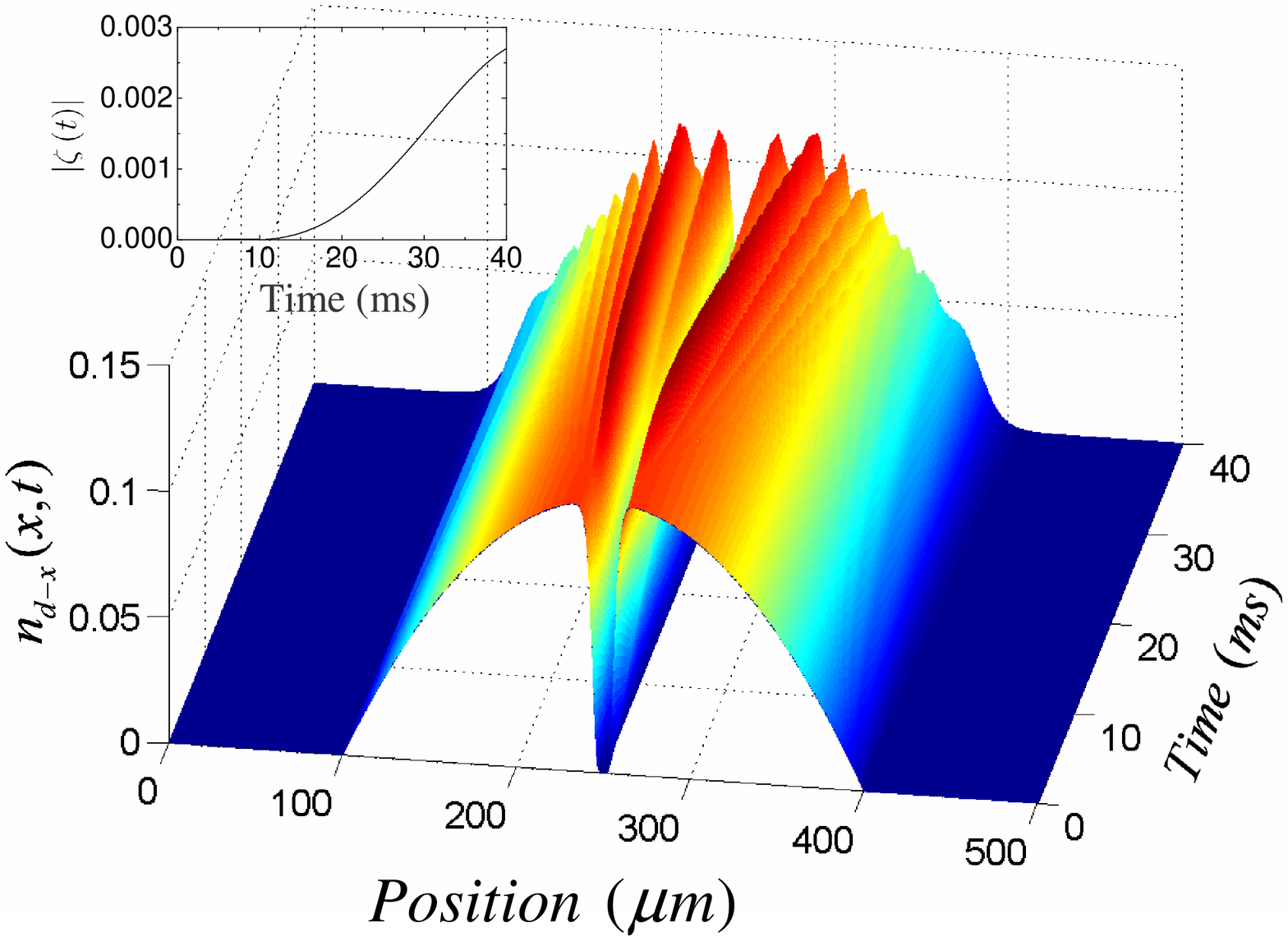}
\caption{Shown is the evolution of the density expectation value calculated
using the experimental parameters in Ref. \protect\cite{Andrew}. It is shown
that there is a clear interference pattern after the overlapping of two
independent interacting condensates. Shown in the inset is the evolution of $%
\left\vert \protect\zeta \left( t\right) \right\vert $ for the experimental
parameters in Ref. \protect\cite{Andrew}.}
\end{figure}

Based on the evolution of the density expectation value shown in Fig. 2, we
see that the overall width (about $300$ $\mathrm{\mu m}$) of the ultracold
gases in $x-$direction does not increase obviously. This is due to the fact
that the initial density distribution is cigar-shaped, and thus the
expansion in $x-$direction is very slow, while the expansion in $y$ and $z$
directions is much quick. In the experiment of Ref. \cite{cigar}, one can
see clearly that there is no obvious expansion in the long $x-$axis for
cigar-shaped condensate. For two initially independent condensates, when the
double-well potential is switched off, one should note that in the regime
close to $x=250$ $\mathrm{\mu m}$ shown in Fig. 2, the ultracold gases
expand rapidly in $x-$direction because in this regime the ultracold gases
have higher kinetic energy. Thus, although the total width of the system in $%
x-$direction does not increase obviously, the rapid expansion in the central
regime will lead to the overlapping between two initially independent
condensates, and results in the interference effect. The overall width of
the system shown in Fig. 2 is smaller than the experimental result of about $%
500$ $\mathrm{\mu m}$ \cite{Andrew}. This difference may come from the
expansion of thermal cloud in this experiment \cite{Andrew}. After $40$
\textrm{ms} expansion, the numerical result in \cite{Rohrl} for two
coherently separated condensates also showed that the overall width of the
system is about $300$ $\mathrm{\mu m}$.

The essential reason for the emergence of the interference term of two
initially independent condensates lies in that because of the exchange
symmetry of identical bosons and interatomic interaction, the two initially
independent condensates will become coherent after the overlapping between
the two condensates. The physical mechanism of this interaction-induced
coherence can be also understood based on the second quantization method.
For two initially independent condensates, the quantum state of the whole
system can be written as \cite{Pethick}
\begin{equation}
\left\vert N_{1},N_{2}\right\rangle =\frac{C_{n}}{\sqrt{N_{1}!N_{2}!}}(%
\widehat{a}_{1}^{\dag })^{N_{1}}(\widehat{a}_{2}^{\dag })^{N_{2}}\left\vert
0\right\rangle ,  \label{initial-state}
\end{equation}%
where $C_{n}$ is a normalization constant to assure $\left\langle
N_{1},N_{2}|N_{1},N_{2}\right\rangle =1$. $\widehat{a}_{1}^{\dag }$ ($%
\widehat{a}_{2}^{\dag }$) is a creation operator which creates a particle
described by the single-particle state $\phi _{1}$ ($\phi _{2}$) in the left
(right) condensate. The density expectation value is then%
\begin{equation}
n_{d}\left( \mathbf{r},t\right) =\left\langle N_{1},N_{2},t\right\vert
\widehat{\Psi }^{\dag }\widehat{\Psi }\left\vert N_{1},N_{2},t\right\rangle .
\label{density-second}
\end{equation}

Here $\widehat{\Psi }\left( \mathbf{x},t\right) $ is the field operator. The
operators $\widehat{a}_{1}$ and $\widehat{a}_{2}$ can be written as $%
\widehat{a}_{1}=\int \widehat{\Psi }\phi _{1}^{\ast }dV$ and $\widehat{a}%
_{2}=\int \widehat{\Psi }\phi _{2}^{\ast }dV$, respectively. By using the
commutation relations of the field operators $[\widehat{\Psi }\left( \mathbf{%
x},t\right) ,\widehat{\Psi }\left( \mathbf{y},t\right) ]=0$ and $[\widehat{%
\Psi }\left( \mathbf{x},t\right) ,\widehat{\Psi }^{\dagger }\left( \mathbf{y}%
,t\right) ]=\delta \left( \mathbf{x}-\mathbf{y}\right) $. It is easy to get
the commutation relation $[\widehat{a}_{1},\widehat{a}_{2}^{\dagger }]=\zeta
^{\ast }$. We see that $\widehat{a}_{1}$ and $\widehat{a}_{2}^{\dagger }$
are not commutative any more for $\int \phi _{1}\phi _{2}^{\ast }dV$ being a
nonzero value. The non-commutative property between $\widehat{a}_{1}$ and $%
\widehat{a}_{2}^{\dagger }$ plays an essential role in the emergence of the
interference term. Our calculations show that the result given by Eq. (\ref%
{density-second}) is the same as the result given by Eq. (\ref{density})
which is obtained based on the many-body wavefunction. We have also proven
that the evolution equations for $\phi _{1}$ and $\phi _{2}$ based on the
many-body wavefunction are the same as the results based on the second
quantization method \cite{Xiong}.

One can get a further physical picture for the emergence of interference
pattern of two initially independent condensates through a general
investigation for the case of $N_{1}\left\vert \zeta \right\vert >>1$, $%
N_{2}\left\vert \zeta \right\vert >>1$ and $N_{1}\sim N_{2}$. Based on our
numerical calculations, these conditions are satisfied for the case of $40$
\textrm{ms} expansion in the experiment of Ref. \cite{Andrew}. When these
conditions are satisfied, $NB_{n}a_{d}\approx N_{1}$, $NB_{n}b_{d}\approx
\sqrt{N_{1}N_{2}}$ and $NB_{n}c_{d}\approx N_{2}$. Thus, the density
expectation value can be approximated very well as
\begin{equation}
n_{d}\left( \mathbf{r},t\right) \simeq \Phi _{e}^{\ast }\left( \mathbf{r}%
,t\right) \Phi _{e}\left( \mathbf{r},t\right) ,  \label{factor-com}
\end{equation}%
where the effective order parameter $\Phi _{e}\left( \mathbf{r},t\right) $
is
\begin{equation}
\Phi _{e}\left( \mathbf{r},t\right) =\sqrt{N_{1}}\phi _{1}\left( \mathbf{r}%
,t\right) +\sqrt{N_{2}}e^{i\varphi _{c}}\phi _{2}\left( \mathbf{r},t\right) .
\label{effective-order}
\end{equation}%
In particular, after removing the double-well potential, the evolution of
the effective order parameter $\Phi _{e}\left( \mathbf{r},t\right) $ can be
described very well by the following Gross-Pitaevskii equation \cite{Xiong}%
\begin{equation}
i\hslash \frac{\partial \Phi _{e}}{\partial t}\simeq -\frac{\hslash ^{2}}{2m}%
\nabla ^{2}\Phi _{e}+g\left\vert \Phi _{e}\right\vert ^{2}\Phi _{e}.
\label{app-equation}
\end{equation}%
We see that the emergence of the effective order parameter gives us strong
evidence that the coherence is formed between two initially independent
condensates, and thus results in the emergence of high-contrast interference
fringes.

In summary, upon expansion, we calculate the density expectation value of
two initially independent condensates. Quite different from the current
beliefs, it is found that there is a nonzero interference term when the
interatomic interaction and the exchange symmetry of identical bosons are
both considered carefully. In fact, it is well-known that the interaction
plays an essential role in the formation of the order parameter of
Bose-condensed gases, i.e. the formation of a stable coherent property.
Here, we provide an example in which the interaction induces coherent
evolution between two initially independent condensates. To investigate more
clearly the interference pattern due to the measurement process alone, an
experimental investigation of two independent ideal condensates would be
very interesting, because there is no interference term for ideal
condensates in the density expectation value. In the last few years, the
rapid experimental advances of Feshbach resonance where the scattering
length can be tuned from positive to negative make this sort of experiment
be feasible.


\begin{thebibliography}{99}
\bibitem{Xiong} Detailed analyses will be given elsewhere.
\end{thebibliography}
\end{document}